\documentclass[aps,prd,10pt,twocolumn,superscriptaddress,nofootinbib,showkeys,showpacs,altaffilletter]{revtex4-1}

\usepackage{graphicx}
\usepackage{dcolumn}
\usepackage{amssymb}
\usepackage{amsmath}
\usepackage{amsfonts}
\usepackage{amsbsy}
\usepackage{color}
\usepackage{rotating}
\usepackage[english]{babel}
\usepackage{soul}

\usepackage{hyperref}
\hypersetup{
    colorlinks=true,
    linkcolor=blue,
    filecolor=magenta,
    urlcolor=cyan,
    citecolor=cyan
}

\newcommand{\be}{\begin{equation}}
\newcommand{\ee}{\end{equation}}
\newcommand{\bea}{\begin{eqnarray}}
\newcommand{\eea}{\end{eqnarray}}

\begin{document}

\title{Cosmological Constraints on Entropic Cosmology with Matter Creation}

\date{\today}

\author{Hussain Gohar}
\email{hussain.gohar@sns.nust.edu.pk}
\affiliation{Department of Physics, School of Natural Sciences, National University of Sciences and Technology, H-12, 44000 Islamabad, Pakistan}
\author{Vincenzo Salzano}
\email{vincenzo.salzano@usz.edu.pl}
\affiliation{Institute of Physics, University of Szczecin, Wielkopolska 15, 70-451 Szczecin, Poland}

\begin{abstract}
We investigate entropic force cosmological models with the possibility of matter creation and energy exchange between the bulk and the horizon of a homogeneous and an isotropic flat Universe. We consider three different kinds of entropy, Bekenstein's, the non-extensive Tsallis-Cirto's and the quartic entropy, plus some phenomenological functional forms for matter creation rate to model different entropic force models and put the observational constraints on them. We show that while most of them are basically indistinguishable from a standard $\Lambda$CDM scenario, the Bekenstein entropic force model with a matter creation rate proportional to the Hubble parameter is statistically highly favored over $\Lambda$CDM. As a general result, we also find that both the Hawking temperature parameter $\gamma$, which relates the energy exchange between the bulk and the boundary of the Universe, and the matter creation rate $\Gamma(t)$, must be very small in order to reproduce observational data.
\end{abstract}

\maketitle

\section{Introduction}

In the 1970s, Bekenstein and Hawking \cite{Hawking:1974rv,Bekenstein:1973ur} showed that the laws of black hole thermodynamics are similar to the standard thermodynamics \cite{Bardeen:1973gs}. Since then a large number of gravitational and cosmological applications \cite{Jacobson:1995ab,Eling:2006aw,Verlinde:2010hp,Padmanabhan:2003pk,Padmanabhan:2009vy,Padmanabhan:2009kr,Cai:2010hk,Gao:2010fw,Li:2010bc,Casadio:2010fs,Nicolini:2010nb,Modesto:2010rm,Wei:2010am,Cai:2008ys,Cai:2005ra,Cai:2006rs,Sheykhi:2010wm,Easson:2010av,Easson:2010xf,Komatsu:2016vof} have been investigated by employing the correspondence between the area of the event horizon and the surface gravity of the black hole with the entropy and the temperature in standard thermodynamics. Jacobson \cite{Jacobson:1995ab} derived the Einstein field equations from the proportionality of the entropy and the horizon area by assuming the heat flow across the horizon. In this way, Padmanabhan \cite{Padmanabhan:2003pk, Padmanabhan:2009vy} used the holographic equipartition law, which states that the expansion of the cosmic space is due to the difference between the degrees of freedom on the surface and in the bulk of a region of space, to derive the Friedmann and acceleration equations. Furthermore, a radical notion was given by Verlinde \cite{Verlinde:2010hp}:  he defined gravity as an entropic force, which is originated in a system as a result of the statistical tendency to increase its entropy. He used the holographic principle \cite{tHooft:1993dmi}, which states that the microscopic degrees of freedom could be represented holographically on the horizon and these degrees of freedom could be measured in terms of entropy. All these new approaches give a new insight into the problem of quantum gravity, which could possibly explore the emergence of space-time from a thermodynamic perspective.

Current observational data \cite{Perlmutter:1998np,Riess:1998cb,Riess:2004nr,Riess:2006fw,Scolnic:2017caz,Hinshaw:2012aka,Aghanim:2018eyx,Alam:2020sor} show that the expansion of the Universe is accelerating and a large number of investigations have been done to understand this experimental fact \cite{Li:2011sd,Bamba:2012cp}. The standard approach to explain the accelerated expansion of the Universe is to add and include in the cosmic inventory a new component, the so called dark energy fluid, with exotic properties with respect to other standard energy-matter contribution, and whose effects into the Einstein field equations are exactly those to lead the acceleration. Despite many successes, dark energy models are not able to explain in a satisfactory way all the observational probes we have collected till now, and are affected by some additional theoretical shortcomings \cite{Bull:2015stt}.

Therefore, many alternative ideas have been investigated to solve and understand the problems \cite{Bamba:2012cp,Li:2011sd,Nojiri:2006ri,Sola:2013gha}. Here, in particular, we will focus on the work from Easson et al. \cite{Easson:2010av,Easson:2010xf}, who gave a new perspective to explain the accelerated expansion: entropic cosmology. According to the entropic cosmology approach, additional entropic force terms are needed to be added to the Einstein field equations, and are assumed to be coming from the neglected boundary terms in the Einstein Hilbert action. The physical motivation behind the entropic force terms is due to the idea of having information holographically stored at the boundary of the Universe. Hence, associated to the entropy and the temperature on the boundary of the Universe, there might be an entropic force acting on the boundary of the Universe, which could be responsible for
an early and the late accelerated expansion of the Universe.

In original entropic force models \cite{Easson:2010av,Easson:2010xf}, the Hawking temperature \cite{Hawking:1974rv} and the Bekenstein entropy \cite{Bekenstein:1973ur} are being used. By using the holographic principle \cite{tHooft:1993dmi} and considering the Hubble horizon as the boundary of the homogeneous and an isotropic Universe, an associated entropy and a temperature are being defined on the Hubble horizon, which extend the Bekenstein entropy and the Hawking temperature for the case of the entire Universe. The Bekenstein entropy $S$ and the Hawking temperature $T$ on the Hubble horizon $r_H$, are given by
\be\label{entropy1}
S=\frac{k_Bc^3A}{4\hbar G},~~~~~~T=\gamma \frac{\hbar c}{2\pi k_B r_H},
\ee
where $r_H$ and the surface area $A$ of the sphere with the Hubble horizon are defined as
\be\label{area}
r_H=\frac{c}{H}, ~~~~~~ A=4\pi r_H^2.
\ee
Here, $H=\dot a/a$ is the Hubble parameter and $a(t)$ is the scale factor at a time $t$; $G$, $c$, $\hbar$ and $k_B$ are respectively the Newton's gravitational constant, the speed of light, the reduced Planck's constant and the Boltzmann's constant; $\gamma$ is a non negative free parameter assumed to be order of one by theoretical considerations \cite{Easson:2010av,Komatsu:2013qia}.

The entropic force $F_r$ on the Hubble horizon can be defined as
\be
F_r=-\frac{dE}{dr_H}=-T\frac{dS}{dr_H}, \label{entropicforce}
\ee
where the minus sign shows the direction of increasing entropy. By using Eqs.~(\ref{entropy1}) and (\ref{area}) in Eq.~(\ref{entropicforce}), we have the entropic force on the horizon
\be
F_r=-\gamma\frac{c^4}{G}, \label{entropicforce1}
\ee
and this force is assumed to be responsible for the accelerated expansion of the Universe. It is interesting to note that, for $\gamma=1/4$, the entropic force $F_r$ becomes the maximum force in general relativity \cite{Gibbons:2002iv,Barrow:2014cga,Dabrowski:2015eea,Ong:2018xna}.

Similarly, the entropic pressure $p_e$ on the Hubble horizon due to the entropic force $F_r$ can be written as
\be
p_e=\frac{F_r}{A}=-\gamma\frac{c^2}{4\pi G}H^2. \label{entropicpressure}
\ee
By considering a homogeneous and an isotropic Universe filled with a perfect fluid having a pressure $p$, we can define the effective pressure $p_{eff}$
\be
p_{eff}=p+p_e. \label{peff1}
\ee
This can be later used in the acceleration and the continuity equations to define the entropic force terms for the entropic force models.

In standard entropic force models, where the Bekenstein entropy and the Hawking temperature are used, the acceleration equation for a homogeneous and an isotropic expanding Universe is given by \cite{Easson:2010av,Easson:2010xf}
\be
\frac{\ddot a}{a}=-\frac{4\pi G}{3}\left(\rho+\frac{3p}{c^2}\right)+\gamma H^2, \label{A1}
\ee
where $\rho$ represents the total energy density of the Universe. The corresponding Friedmann equation for the  phenomenological entropic force model can be written as
\be
H^2=\frac{8\pi G}{3}\rho+\gamma H^2. \label{F1}
\ee
Here, $\gamma H^2$ is the entropic force term which could be alternative to the cosmological constant $\Lambda$ in the standard $\Lambda$CDM cosmology. The continuity equation can be derived from the first law of thermodynamics, and is given by
\be
\dot \rho+3H\left(\rho+\frac{p}{c^2}\right)=-\gamma\frac{3}{4\pi G}H\dot H. \label{ceq1}
\ee
In \cite{Easson:2010av,Easson:2010xf}, the right hand side of the equation is zero, based on the assumptions of adiabatic processes, where the entropy of the universe remains constant. But  the entropy of the Universe must be increasing, and by considering the non-adiabatic processes across the horizon, we have the nonzero right hand side in Eq.~(\ref{ceq1}).

Another way to introduce the entropic force terms is by using the effective pressure $p_{eff}$, given in Eq.~(\ref{peff1}). By introducing the effective pressure in the acceleration and continuity equations, we have
\be
H^2=\frac{8\pi G}{3}\rho, \label{F11}
\ee
\be
\frac{\ddot a}{a}=-\frac{4\pi G}{3}\left(\rho+\frac{3p_{eff}}{c^2}\right), \label{A11}
\ee
\be
\dot \rho+3H(\rho+\frac{p_{eff}}{c^2})=0. \label{C11}
\ee
By using Eq.~(\ref{peff1}), the above equations can be written as
\be
H^2=\frac{8\pi G}{3}\rho, \label{F2}
\ee
\be
\frac{\ddot a}{a}=-\frac{4\pi G}{3}\left(\rho+\frac{3p}{c^2}\right)+\gamma H^2, \label{A2}
\ee
\be
\dot \rho+3H\left(\rho+\frac{p}{c^2}\right)=\gamma \frac{3}{4\pi G}H^3. \label{C2}
\ee
In the original entropic force models \cite{Easson:2010av,Easson:2010xf}, only $H^2$ term, or the combination of $H^2$ and $\dot H$ are included in the Friedmann and acceleration equations as extra driving terms, coming from the usually neglected boundary terms. The problem with considering only $H^2$ term is that these models do not describe a decelerating and an accelerating Universe \cite{Basilakos:2014tha,Basilakos:2012ra,Gomez-Valent:2014fda}. It has been pointed out by Basilakos et al. \cite{Basilakos:2014tha} that the original entropic force model does not explain the cosmological fluctuations and are not consistent with the structure formation. However, by including a term $H$, the modified entropic force models can describe a decelerating and an accelerating Universe. But for structure formation, both $H$ and a constant entropic force terms are needed to tackle the problem. In principle, there can be higher orders terms like $H^4$ in modified entropic models, which are needed to discuss the entropic inflation \cite{Easson:2010xf}. In this paper, we are not interested in models of inflation in entropic cosmology \cite{Qiu:2011zr,Cai:2010kp,Cai:2010zw}.

It is pertinent to mention here that the entropic force models are completely different from the Verlinde's entropic gravity, where the gravity is itself an entropic force. For more details see \cite{Verlinde:2010hp, Visser:2011jp,Kobakhidze:2010mn}.

The form of the driving entropic force terms depend on the choice of the definition of the entropy. Komatsu and Kimura \cite{Komatsu:2013qia,Komatsu:2014ywa,Komatsu:2015nga,Komatsu:2015nkb,Komatsu:2013qia,Komatsu:2014vna,Komatsu:2014lsa,Komatsu:2012zh} modified the entropic force models by incorporating the non-extensive Tsallis-Cirto entropy \cite{Tsallis:1987eu,Tsallis:2012js} and the quartic entropy \cite{Komatsu:2013qia}.  In original entropic force model \cite{Easson:2010av,Easson:2010xf}, Easson et al. used the Bekenstein entropy, which is proportional to the area on the Hubble horizon, to get the $H^2$ term. Tsallis and Cirto introduced a non-extensive entropy for black holes and Komatsu and Kimura applied the Tsallis-Cirto entropy, which is proportional to the volume of the Hubble horizon, to the cosmological horizons. In \cite{Komatsu:2013qia}, it is applied to the Hubble horizon to get the $H$ term in the modified entropic model. A higher dimensional entropy, called the quartic entropy, is also being used to discuss the constant entropic force term in the modified entropic force models. Hence, the modified quartic entropic force model tackles the problem of a decelerating and an accelerating Universe and the problem of structure formation. A general formalism for all the entropic force models have been discussed in \cite{Komatsu:2015nkb}.

In \cite{Komatsu:2014lsa}, Komatsu and Kimura have categorized the entropic force models into two categories: the varying $\Lambda (t)$ type and the Bulk viscous ($BV$) type models. In $\Lambda (t)$ type models, the driving entropic force terms are added to both Friedmann and acceleration equations. The continuity equation has non zero right hand side and these types of entropic force models are similar to energy exchange cosmological models \cite{Barrow:2006hia}, where the energy exchange between two cosmological fluids are considered like, for example, the interaction between the dark energy and the dark matter \cite{Wang:2005jx,Wang:2007ak,Pavon:2005yx,Setare:2006wh,Hu:2006ar} or the coupling between the matter and the radiation \cite{Szydlowski:2005ph}. The original entropic force model in \cite{Easson:2010av,Easson:2010xf} is  of $\Lambda (t)$ type because in Eqs.~(\ref{A1}) and (\ref {F1}), we have the $\gamma H^2$ in both Friedmann and acceleration equations. On the other hand, for the case of $BV$ type models, which are inspired by the possible creation of cold dark matter \cite{Zimdahl:2000zm,Jesus:2011ek,Lima:2012cm,Lima:2009ic,Freaza:2002ic} and bulk viscosity of the cosmological fluids \cite{Murphy:1973zz,Barrow:1986yf,Barrow:1988yc,Davies:1987ti,Lima:1987fj,Zimdahl:1996ka,Brevik:2005bj,Ren:2005nw,Capozziello:2005pa,Fabris:2005ts,Colistete:2007xi,Avelino:2008ph,HipolitoRicaldi:2009je,Avelino:2010pb,Piattella:2011bs,Lima:1995xz,Lima:2008qy}, the driving entropic force terms are only included in the acceleration equation but not in the Friedmann equation due to dissipation processes. The $BV$ type models can be interpreted as bulk viscous cosmological models, where the  bulk viscosity generates the entropy in an isotropic and a homogeneous Universe.
One can see Eqs.~(\ref{A11}) and (\ref{C11}) \cite{Komatsu:2014lsa}, where $p_{eff}=p+p_e$ is introduced in the acceleration and continuity equations to get the required driving entropic force terms, whereas there is not any entropic force term in the Friedmann equation Eq.~(\ref{F11}).

\section{General Formalism for Entropic Force Models }

By following \cite{Komatsu:2015nkb}, we consider a homogeneous and an isotropic Universe with the energy exchange between the bulk and the Hubble horizon. In addition to this, we also consider the possible creation of matter in the Universe. These models will be a combination of both $\Lambda (t)$ type and $BV$ type entropic force models. The energy exchange is related to the reversible entropy; while the irreversible entropy will be with the creation of matter in the Universe. We take into account the general form of entropy given in \cite{Komatsu:2015nkb} for the case of reversible entropy defined on the Hubble horizon, which can be written as
\be
S_m=\frac{\pi k_B c^3}{\hbar G}L_mr^m_H \label{entgen}
\ee
where $m$ takes the values $2$, $3$ and $4$, which correspond to Bekenstein, Tsallis-Cirto and quartic entropies.  $L_m$ is a free parameter and $L_2=1$ for $m=2$, $L_3=\xi$ for $m=3$ and $L_4=\chi$  for $m=4$ respectively. Here, $\xi$ and $\chi$ are also nonnegative parameters correspond to Tsallis-Cirto entropy and the quartic entropy. A detailed study of modified entropic force models related to these entropies have been investigated in \cite{Komatsu:2013qia}.

The entropic force by using the general form of the entropy can be derived by using Eq.~(\ref{entgen}), which can be written as
\be
F_{m}=-T\frac{dS_{m}}{dr_H}=-\gamma \frac{c^4}{G}\left(\frac{mL_m}{2}\right)r_H^{m-2},
\ee
and from it, the general expression for the entropic pressure $p_{rm}$ due to the general form of the entropic force $F_{rm}$ on the boundary of the Universe can be written as
\be
p_{m}=\frac{F_{m}}{A}=-\gamma \left(\frac{c^mmL_m}{8\pi G}\right)H^{4-m}.
\ee
The Friedmann and acceleration equations for the general entropic force models are given in \cite{Komatsu:2015nkb}, and can be written as
\be
H^2=\frac{8\pi G}{3}\rho+\gamma\left(\frac{c^{m-2}mL_m}{2}\right)H^{4-m}, \label{friedeq}
\ee
\be
\frac{\ddot a}{a}=-\frac{4\pi G}{3}\left(\rho+\frac{3p_{eff}}{c^2}\right)+\gamma\left(\frac{c^{m-2}mL_m}{2}\right)H^{4-m}. \label{acceq}
\ee
By using Eqs.~(\ref{friedeq}) and (\ref{acceq}), we have the continuity equation
\be
\dot\rho+3H\left(\rho+\frac{p_{eff}}{c^2}\right)=
-\gamma\left(\frac{3c^{m-2}mL_m}{8\pi G}\right)\left(\frac{4-m}{2}\right)H^{3-m}\dot H, \label{conteq}
\ee
where the effective pressure $p_{eff}$ reads as
\be
p_{eff}=p+p_{irr}=p-\frac{(\rho c^2+p)\Gamma(t)}{3H}. \label{peff}
\ee
Here, the function $\Gamma(t)$ represents the particle production rate and it is related with the creation pressure $p_{irr}$. It is completely unknown and yet to be defined in quantum field theory \cite{Parker:1968mv,Parker:1969au,Parker:1971pt,Birrell:1979pi,Zeldovich:1977vgo,Prigogine:1989zz}.
For this study, we take the particle production rate or the entropy production function as much general as possible, and we rely on the phenomenological functions of $\Gamma(t) = \Gamma_0, \Gamma_0H, \Gamma_0H^2,\Gamma_0/H$ analyzed in \cite{Zimdahl:1996ka,Pan:2018ibu,Paliathanasis:2016dhu,Pan:2016jli,Abramo:1996ip,Gunzig:1997tk,Lima:1992np}.

For this study, we assume an open system  undergoing non-adiabatic processes of matter creation, which generate an irreversible entropy. We also consider the energy exchange between the bulk and the boundary as reversible processes, which corresponds to the reversible entropy. In the above equations, the terms with $\gamma$ correspond to reversible processes and those ones with $p_{irr}$ to irreversible processes.

In order to understand the reversible and irreversible processes related to the energy exchange between the bulk and the horizon and the matter creation in the universe, we review (for detailed calculations, see \cite{Komatsu:2015nkb} and references therein) the first law of thermodynamics for an open system containing $N(t)$ total number of particles in a volume $V(t)$, which can be written as
\be
\frac{d}{dt}(\epsilon V)+p\frac{dV}{dt}=\left(\frac{dQ}{dt}\right)_{rev}+\left(\frac{\epsilon+p}{n}\frac{d}{dt}(nV)\right)_{irr}, \label{Ceq1}
\ee
where $\epsilon=\rho c^2$ is the energy density of the fluid and $n=N/V$ is the particle number density. The first term in the right hand side of Eq.~(\ref{Ceq1}) shows the heat flow $dQ=TdS$ across the horizon and by using the general form of the entropy, Eq.~(\ref{entgen}), at $r=r_H$, we have the expression for $dQ/dt$
\be
\frac{dQ}{dt}=\gamma\frac{c^4}{G}\left(\frac{mL_m}{2}\right)r_H^{m-2}\frac{dr_H}{dt}. \label{Q1}
\ee
The second term in the right hand side of Eq.~(\ref{Ceq1}) is related to the  matter creation, which corresponds to the irreversible entropy, and it  can be written as
\be
\frac{\epsilon+p}{n}\frac{d}{dt}(nV)=\left(\rho+\frac{p}{c^2}\right)c^2V\Gamma(t). \label{Q2}
\ee
where $\dot n+3\frac{\dot a}{a}n=n\Gamma(t)$ has been used to get the above equation. Here,  $\Gamma(t)$ represents the particle production rate.
By using Eqs.~(\ref{Q1}) and (\ref{Q2}) in Eq.~(\ref{Ceq1}), we get the continuity equation from the first law of thermodynamics
\be
\dot\rho+3H\left(\rho+\frac{p_{eff}}{c^2}\right)= -\gamma\left(\frac{3c^{m-2}mL_m}{8\pi G}\right)H^{3-m}\dot H,
\ee
where $p_{eff}$ is given in Eq.~(\ref{peff}).
Note that there is an extra $(4-m)/2$ in the right hand side of the continuity equation Eq.~(\ref{conteq}), and that continuity equation is derived from Friedmann and acceleration equations. In \cite{Komatsu:2015nkb} both the continuity equations have been calculated and compared in detail. Both the continuity equations are equivalent for $m=2$. However, we have used the continuity equation Eq.~(\ref{conteq}) for this study to analyze the observational data.

In the following subsections, we consider more specifically three entropic force models corresponding to Bekenstein, Tsallis-Cirto and the quartic entropy.

\subsection{Bekenstein Entropic Force Model}

For $m=2$, Eqs.~(\ref {friedeq})~-~(\ref{acceq}) and (\ref{conteq}) reduce to:
\be
H^2=\frac{8\pi G}{3}\rho+\gamma H^2 \label{friedeqm1}\, ,
\ee
\be
\frac{\ddot a}{a}=-\frac{4\pi G}{3}\left(\rho+\frac{3p_{eff}}{c^2}\right)+\gamma H^2 \label{acceqm1}\, ,
\ee
\be
\dot\rho+3H\left(\rho+\frac{p_{eff}}{c^2}\right)=-\gamma\frac{3}{4\pi G}H\dot H. \label{conteqm1} \,
\ee
Making explicit a multi-fluid scenario, the Friedmann equation (\ref{friedeqm1}) can be rearranged as
\be
H^2 = \frac{8\pi G}{3 (1-\gamma)} \sum_{i} \rho_{i},\label{friedeqm1a}
\ee
and the corresponding continuity equation as
\begin{equation}
\sum_{i} \dot{\rho}_{i} + 3 H \left[\sum_{i} \left(\rho_{i} + \frac{p_{i}}{c^2}\right) + \frac{p_{irr}}{c^2}\right] = -\gamma \frac{3}{4 \pi G} H \, \dot{H}
\end{equation}
where, as above
\begin{equation}
p_{irr} = -\frac{\Gamma(t)}{3 H} \sum_{i} \left(\rho_{i}c^2 + p_{i}\right).
\end{equation}
The summation over index $i$ runs over the energy density of matter $\rho_m$, of radiation $\rho_r$, and of dark energy which will be, for us, always a cosmological constant $\rho_{\Lambda}$. By using Eq.~(\ref{friedeqm1a}), we have
\begin{equation}
\sum_{i} \dot{\rho}_{i} + 3 (1-\gamma) H \left[\sum_{i} \left(\rho_{i} + \frac{p_{i}}{c^2}\right) + \frac{p_{irr}}{c^2}\right] = 0. \label{coneq2}
\end{equation}
In the following, we use $\Gamma(t)$ functions taken from the literature \cite{Zimdahl:1996ka,Pan:2018ibu,Paliathanasis:2016dhu,Pan:2016jli,Abramo:1996ip,Gunzig:1997tk,Lima:1992np} and rewrite the continuity equations for each case. In order to solve these continuity equations we must define the initial conditions. We choose to set them at $a=1$ (or equivalently, at redshift $z=0$), connecting the present-time densities $\rho_{i,0}$ to the appropriate observable cosmological parameters, the dimensionless density parameters $\Omega_{i}$, by using the relation
\begin{equation}
\rho_{i,0} = \frac{3 H^{2}_{0}}{8 \pi G} \Omega_{i} \; .
\end{equation}
Thus, specifically for the Bekenstein entropic force model, the Friedmann equation Eq.~(\ref{friedeqm1a}) becomes
\begin{equation}
H(a) = H_0 \frac{\sqrt{\Omega_m a^{-3} + \Omega_r a^{-4} + \Omega_{\Lambda}}}{\sqrt{1-\gamma}}\; ,
\end{equation}
with $\Omega_{\Lambda}$ defined by the normalization condition $H(a=1)= H_0$ as
\begin{equation}
\Omega_{\Lambda} = 1 - \Omega_m - \Omega_r - \gamma\; .
\end{equation}
At this point one could argue that one should be able to constrain only the parameters $\Omega'_{i} = \Omega_{i} / (1-\gamma)$, which might look as the only directly measurable ones. But actually, as we will show in the next subsections, we can clearly separate and distinguish the weight of $\gamma$ from that of the $\Omega_{i}$ in the cosmological background evolution, for the way it enters and influence the continuity equations.

\subsubsection{\texorpdfstring{$\Gamma = \Gamma_0 = const.$}{GammaH0}}

By using Eq.~(\ref{friedeqm1a}), and converting from time to scale factor derivative, we can rewrite the continuity equation Eq.~(\ref{coneq2}) as
\bea
&&\rho'_{i} + \frac{3 \left(1+w_i\right)\left(1-\gamma\right)}{a} \rho_{i} -  \nonumber\\
&&\frac{\left(1+w_i\right)\left(1-\gamma\right)\Gamma_{0}}{a} \frac{\rho_{i}}{\kappa_B\left( \gamma\right) \left(\sqrt{\sum_{i}\rho_{i}}\right)  } = 0,
\eea
where prime denotes the derivative with respect to the scale factor $a(t)$. The parameter $\kappa_{B}\left( \gamma\right)$ is given by
\be
\kappa_{B}^2\left( \gamma\right)=\frac{8\pi G}{3 \left( 1-\gamma\right)}.
\ee
Note that the time behaviour of the cosmological constant is still constant in this scenario $(w_{\Lambda}=-1)$, while matter and radiation may behave differently. Moreover, all fluids are coupled.

\subsubsection{\texorpdfstring{$\Gamma(t) = \Gamma_0 H$}{GammaH1}}

The continuity equation Eq.~(\ref{coneq2}) for this case is
\be
\rho'_{i} + \frac{3\left(1+w_i\right)\left(1-\gamma\right) \left(1-\frac{\Gamma_0}{3}\right)}{a} \rho_i = 0.
\ee
Clearly, the fluids are separable and for each one the density behaves as
\be
\rho_i=\rho_{i,0}a^{-\tilde{\Gamma}},
\ee
where $\rho_{i,0}$ is an integration constant (density today) and
\be
\tilde{\Gamma}= 3\left(1+w_i\right)\left(1-\gamma\right) \left(1-\frac{\Gamma_0}{3}\right).
\ee
Once again, it is possible to see that the cosmological constant is still a constant, while matter and radiation may exhibit some change w.r.t. the standard scenario.

\subsubsection{\texorpdfstring{$\Gamma(t) =\Gamma_0 H^2$}{GammaH2}}

For this case, we have rewritten the continuity equation, Eq.~(\ref{coneq2}), as
\bea
&&\rho'_{i} + \frac{3 \left(1+w_i\right)\left(1-\gamma\right)}{a} \rho_{i} -  \nonumber\\
&&\frac{\left(1+w_i\right)\left(1-\gamma\right)\Gamma_{0}}{a} \kappa_{B}\left( \gamma\right) \left(\sqrt{\sum_{i}\rho_{i}}\right)  \rho_{i} = 0.
\eea

\subsubsection{\texorpdfstring{$\Gamma(t) = \frac{\Gamma_0}{H}$}{GammaHM1}}

Finally, for this case, Eq.~(\ref{coneq2}) read
\bea
&&\rho'_{i} + \frac{3 \left(1+w_i\right)\left(1-\gamma\right)}{a} \rho_{i} - \nonumber\\
&&\frac{\left(1+w_i\right)\left(1-\gamma\right)\Gamma_{0}}{a} \frac{\rho_{i}}{\kappa_{B}^2\left( \gamma\right) \left(\sum_{i}\rho_{i}\right)  } = 0.
\eea

\subsection{Tsallis-Cirto Entropic Force Model}

For $m=3$, Eqs.~(\ref {friedeq})~-~(\ref{acceq}) and (\ref{conteq}) reduce to:
\be
H^2=\frac{8\pi G}{3}\rho+ \left(\frac{3c\xi}{2}\right)\gamma H, \label{friedeqmTC}
\ee
\be
\frac{\ddot a}{a}=-\frac{4\pi G}{3}\left(\rho+\frac{3p_{eff}}{c^2}\right)+ \left(\frac{3c\xi}{2}\right)\gamma H, \label{acceqm2}
\ee
\be
\dot\rho+3H\left(\rho+\frac{p_{eff}}{c^2}\right)=-\left(\frac{3}{8\pi G}\frac{3c\xi}{2}\right)\gamma \dot H. \label{conteqm2}
\ee
First of all, we define $\tilde{\gamma}= 3c\xi/2 \gamma$, not only to improve legibility, but also because, given such a combination, our analysis will be able to put constraints only on $\tilde{\gamma}$ and not on the single parameters $\xi$ and $\gamma$. After that, we note that Eq.~(\ref{friedeqmTC}) is a quadratic equation in $H$. If we solve it, we obtain the only physically well motivated solution
\begin{equation}
H(a) = \frac{1}{2} \left( \tilde{\gamma} + \sqrt{4 H^{2}_{0} \left( \Omega_m a^{-3} + \Omega_r a^{-4} + \Omega_{\Lambda} + \tilde{\gamma}^{2} \right)} \right)\, ,
\end{equation}
with $\Omega_{\Lambda}$ given by
\begin{equation}
\Omega_{\Lambda} = \frac{H_0 \left( 1- \Omega_m -\Omega_r - \tilde{\gamma}\right)}{H_0}\; ,
\end{equation}
once the normalization condition $H(a=1)= H_0$ is applied.

\subsubsection{\texorpdfstring{$\Gamma = \Gamma_0 = const.$}{GammaH0}}

The continuity equation (\ref{coneq2}) becomes
\bea
&&\rho'_{i} \left[ 1+ \frac{\tilde{\gamma}}{\sqrt{\tilde{\gamma}^2 + 4 \kappa^2 \left( \sum_{i}\rho_{i}\right) }}\right] + \frac{3 \left(1+w_i\right)}{a} \rho_{i} -  \nonumber\\
&&\frac{\left(1+w_i\right)\Gamma_0}{a} \frac{\rho_{i}} {\frac{1}{2} \left( \tilde{\gamma} + \sqrt{\tilde{\gamma}^{2} + 4 \kappa^{2} \sum_{i}\rho_{i}} \right)} = 0,
\eea
where now $\kappa^2 = 8 \pi G /3$. Once again, note that the cosmological constant is still a constant, but now all fluids are coupled.

\subsubsection{\texorpdfstring{$\Gamma(t) = \Gamma_0 H$}{GammaH1}}

In this case the continuity equation (\ref{coneq2}) reads
\bea
&&\rho'_{i} \left[ 1+ \frac{\tilde{\gamma}}{\sqrt{\tilde{\gamma}^2 + 4 \kappa^2 \left( \sum_{i}\rho_{i}\right) }}\right] +  \nonumber \\
&&\frac{3}{a} \left(1+w_i\right) \left(1- \frac{\Gamma_0}{3} \right) \rho_{i} = 0\, .
\eea
Once again, as in the previous similar case, the fluids are perfectly separable.

\subsubsection{\texorpdfstring{$\Gamma(t) =\Gamma_0 H^2$}{GammaH2}}

Here, the continuity equation Eq.~(\ref{coneq2}) is
\bea
&&\rho'_{i} \left[ 1+ \frac{\tilde{\gamma}}{\sqrt{\tilde{\gamma}^2 + 4 \kappa^2 \left( \sum_{i}\rho_{i}\right) }}\right] + \frac{3 \left(1+w_i\right)}{a} \rho_{i} -  \nonumber \\
&&\frac{\left(1+w_i\right)\Gamma_0}{a} \rho_{i} \left[\frac{1}{2} \left( \tilde{\gamma} + \sqrt{\tilde{\gamma}^{2} + 4 \kappa^{2} \sum_{i}\rho_{i}} \right)\right] = 0\;.
\eea

\subsubsection{\texorpdfstring{$\Gamma(t) = \frac{\Gamma_0}{H}$}{GammaHM1}}

For this last case, the continuity equation (\ref{coneq2}) is
\bea
&&\rho'_{i} \left[ 1+ \frac{\tilde{\gamma}}{\sqrt{\tilde{\gamma}^2 + 4 \kappa^2 \left( \sum_{i}\rho_{i}\right) }}\right] + \frac{3 \left(1+w_i\right)}{a} \rho_{i} -  \nonumber\\
&&\frac{\left(1+w_i\right)\Gamma_0}{a} \frac{\rho_{i}} {\left[\frac{1}{2} \left( \tilde{\gamma} + \sqrt{\tilde{\gamma}^{2} + 4 \kappa^{2} \sum_{i}\rho_{i}} \right)\right]^{2}} = 0.
\eea

\subsection{Quartic Entropic Force Model}

For $m=4$, Eqs.~(\ref {friedeq}), (\ref{acceq}) and (\ref{conteq}), reduce to:
\be
H^2=\frac{8\pi G}{3}\rho+ 2\gamma c^2\chi \label{friedeqm2},
\ee
\be
\frac{\ddot a}{a}=-\frac{4\pi G}{3}\left(\rho+\frac{3p_{eff}}{c^2}\right)+2\gamma c^2\chi \label{acceqm2},
\ee
\be
\dot\rho+3H\left(\rho+\frac{p_{eff}}{c^2}\right)=0. \label{conteqm2}
\ee
The quartic entropic force model is equivalent to the standard $\Lambda$CDM model by taking the constant entropic force term  as an effective cosmological constant in the Friedmann equation, Eq.~(\ref{friedeqm2}), and in the acceleration equation, Eq.~(\ref{acceqm2}). While from the theoretical point of view there is a huge difference, from the observational one, the two models would be indistinguishable. This is why, we will not consider this model in the following analysis.

\section{Data and Statistics}

The above mentioned entropic cosmological models are going to be compared with the most updated (to our knowledge) set of geometrical cosmological data, namely: Type Ia Supernovae (SNeIa) from the Pantheon sample; Cosmic Chronometers (CC); the gravitational lensing data from COSMOGRAIL's Wellspring project (H0LiCOW); the ``Mayflower'' sample of Gamma Ray Bursts (GRBs); Baryon Acoustic Oscillations (BAO); and Cosmic Microwave Background radiation (CMB) from \textit{Planck} $2018$.

We perform our statistical analysis on two different data set: for the one which we call ``full'', we join early- (CMB and BAO data from SDSS) and late-time observations (SNeIa, CC, H0LiCOW, GRBs and BAO data from WiggleZ); for that one which we call ``late-time'', we only include late-time data. This choice was dictated by recognizing that if on one side it is acknowledged that early-time data are more decisive in constraining cosmological models than late-time ones, mainly due to their higher precision, on the other one they also seem to be statistically biased toward a cosmological constant as dark energy, in an Occam-razor sense. Our hope is that by comparing results from data related to such different epochs, we could have some more neat evidence, if any, into a possible presence of a time varying dark energy candidate and/or an alternative theoretical approach.

The total $\chi^2$ we use in the following is of course defined as the sum of the contributions from each probe. In order to minimize the $\chi^2$ we use our own implementation of a Monte Carlo Markov Chain (MCMC) \cite{Berg,MacKay,Neal} and we test its convergence using the method of \cite{Dunkley:2004sv}. We also try to establish a statistical hierarchy or preference among our models by using the Bayesian Evidence, $\mathcal{E}$, calculated using the nested sampling algorithm described in \citep{Mukherjee:2005wg}. In this case, our reference model is the standard $\Lambda$CDM model, analyzed with the same set of data. We calculate the Bayesian Evidence using the algorithm from. The chosen priors are all uninformative, flat, and as much general and wide as possible, so that any prior dependence \citep{Nesseris:2012cq} of $\mathcal{E}$ is negligible.

The priors we have chosen are basically: $0<\Omega_b < \Omega_m < 1$; $0<h<1$, with $h= H_0/100$; $\gamma > 0$; while no prior on the sign of $\Gamma_0$ has been put, as both of them have intrinsic physical meaning ($>0$ for matter creation; $<0$ for matter annihilation). Actually, given the very small values assumed by $\gamma$ and $\Gamma_0$ we have better worked and constrained their logarithmic versions, $\log \gamma$ and $\log \Gamma_0$, except for the cases with $\Gamma(t)= \Gamma_0\,H$, where we have left $\Gamma_0$. Note also that because of this choice, we had to analyze the cases $\Gamma_0<0$ and $\Gamma_0>0$ separately. Moreover, we have enforced the further controls: that $\Omega_{m,r,\Lambda}>0$ for the full time extension, i.e. $a \in [0,1]$; and that $\Omega_{m,r}>0$ are decreasing functions of time. The only cases where $\Gamma_0<0$ is not shown in the following tables, are those for which, in order to be consistent with data, it must assume values much lower than the required minimal numerical precision.

Eventually, we derive the Bayes Factor as the ratio of evidence between two models, $M_{i}$ and $M_{j}$, $\mathcal{B}^{i}_{j} = \mathcal{E}_{i}/\mathcal{E}_{j}$. Generally speaking, if $\mathcal{B}^{i}_{j} > 1$ than the model $M_i$ is preferred over $M_j$, given the data (we will follow Jeffreys' scale \cite{Jeffreys:1939xee}). Once again, the $\Lambda$CDM model will play the role of the reference models $M_j$.

\subsection{Type Ia Supernovae}

The Pantheon compilation \cite{Scolnic:2017caz} is made of $1048$ objects observed in the redshift range $0.01<z<2.26$. The $\chi^2_{SN}$ is
\begin{equation}
\chi^2_{SN} = \Delta \boldsymbol{\mathcal{\mu}}^{SN} \; \cdot \; \mathbf{C}^{-1}_{SN} \; \cdot \; \Delta  \boldsymbol{\mathcal{\mu}}^{SN} \;,
\end{equation}
where $\Delta\boldsymbol{\mathcal{\mu}} = \mathcal{\mu}_{\rm theo} - \mathcal{\mu}_{\rm obs}$ is the difference between the theoretical and the observed distance modulus for each SNeIa and $\mathbf{C}_{SN}$ is the total covariance matrix. The distance modulus is
\begin{equation}
\mu(z,\boldsymbol{p}) = 5 \log_{10} [ d_{L}(z, \boldsymbol{p}) ] +\mu_0 \; ,
\end{equation}
with
\begin{equation}
d_L(z,\boldsymbol{p})=(1+z)\int_{0}^{z}\frac{dz'}{E(z',\boldsymbol{p})} \,
\end{equation}
being the dimensionless luminosity distance; with $\boldsymbol{\theta}$ we indicate the vector of cosmological parameters on which each model depend. We marginalize the $\chi^{2}_{SN}$ over $\mu_0$ (because of the well known degeneracy between the Hubble constant $H_0$ and the SNeIa absolute magnitude) following \cite{conley}, using
\begin{equation}\label{eq:chis}
\chi^2_{SN}=a+\log \left(\frac{d}{2\pi}\right)-\frac{b^2}{d},
\end{equation}
where $a\equiv\left(\Delta \boldsymbol{\mathcal{\mu}}_{SN}\right)^T \; \cdot \; \mathbf{C}^{-1}_{SN} \; \cdot \; \Delta  \boldsymbol{\mathcal{\mu}}_{SN}$, $b\equiv\left(\Delta \boldsymbol{\mathcal{\mu}}^{SN}\right)^T \; \cdot \; \mathbf{C}^{-1}_{SN} \; \cdot \; \boldsymbol{1}$, $d\equiv\boldsymbol{1}\; \cdot \; \mathbf{C}^{-1}_{SN} \; \cdot \;\boldsymbol{1}$ and $\boldsymbol{1}$ is the identity matrix.

\subsection{Cosmic Chronometers}

Early-Type galaxies undergoing passive evolution and exhibiting characteristic peculiar features in their spectra have been deemed able to provide measurements of the Hubble parameter $H(z)$ \cite{Jimenez:2001gg,Moresco:2010wh}, and are for that called Cosmic Chronometers. We use data from \cite{moresco} from the redshift range $0<z<1.97$. The $\chi^2_{H}$ is defined as
\begin{equation}\label{eq:hubble_data}
\chi^2_{H}= \sum_{i=1}^{24} \frac{\left( H(z_{i},\boldsymbol{p})-H_{obs}(z_{i}) \right)^{2}}{\sigma^2_{H}(z_{i})} \; ,
\end{equation}
where $\sigma_{H}(z_{i})$ are the observational errors on the measured values $H_{obs}(z_{i})$.

\subsection{H0LiCOW}

The main goal of the H0LiCOW collaboration \cite{Suyu:2016qxx} was to use the sensitivity of strong gravitational lensing events to constrain $H_0$. For that purpose $6$ lensed quasars were selected \cite{Wong:2019kwg} for which it was possible to collect multiple images. From them, one could exploit lensing time delay as a cosmological probes; in fact, this quantity depends on the cosmological background, given the expression
\begin{equation}\label{eq:timedelGO}
    t(\boldsymbol{\theta},\boldsymbol{\beta})=\frac{1+z_L}{c}\frac{D_L D_S}{D_{LS}}\left[\frac{1}{2}(\boldsymbol{\theta}-\boldsymbol{\beta})^2-
    \hat{\Psi}(\boldsymbol{\theta})\right].
\end{equation}
where, assuming a standard gravitational lensing configuration \cite{gralen.boo}: $z_L$ is the lens redshift; $\boldsymbol{\theta}$ is the angular position of the image; $\boldsymbol{\beta}$ is the angular position of the source; and $\hat{\Psi}$ is the effective lens potential. The distances $D_S$, $D_L$ and $D_{LS}$ are, respectively, the angular diameter distances from the observer to the source, to the lens, and between source and lens. The angular diameter distance is defined as
\begin{equation} \label{eq:ang_dist}
D_{A}(z,\boldsymbol{p})=\frac{1}{1+z}\int_{0}^{z} \frac{c\, \mathrm{d}z'}{H(z',\boldsymbol{p})} \;,
\end{equation}
from which: $D_{S} = D_{A}(z_S)$, $D_{L} = D_{A}(z_L)$, and $D_{LS} = 1/(1+z_S) \left[(1+z_S)D_S - (1+z_L)D_L\right]$ \cite{Hogg:1999ad}. The scaling factor involving all the cosmological distances, which appears in r.h.s. of Eq.~(\ref{eq:timedelGO}),
\begin{equation}
D_{\Delta t} \equiv (1+z_L)\frac{D_L D_S}{D_{LS}}\, ,
\end{equation}
is called time-delay distance and is constrained by H0LiCOW. The data ($D^{obs}_{\Delta t,i}$) and the corresponding errors $(\sigma_{D_{\Delta t,i}})$ for the $6$ quasars are given in \cite{Wong:2019kwg}. Finally, the $\chi^2$ for H0LiCOW data is
\begin{equation}\label{eq:cow_data}
\chi^2_{HCOW}= \sum_{i=1}^{6} \frac{\left( D_{\Delta t,i}(\boldsymbol{p})-D^{obs}_{\Delta t,i}\right)^{2}}{\sigma^2_{D_{\Delta t,i}}} \; ,
\end{equation}

\subsection{Gamma Ray Bursts}

We work with the ``Mayflower'' sample, $79$ GRBs in the redshift interval $1.44<z<8.1$ described in \cite{Liu:2014vda}, which have been calibrated with a robust cosmological model independent procedure. The main GRBs observable is the distance modulus, so the same procedure used for SNeIa is also applied here. The $\chi_{G}^2$ is thus given by $\chi^2_{GRB}=a+\log d/(2\pi)-b^2/d$ as well, with $a\equiv \left(\Delta\boldsymbol{\mathcal{\mu}}^{G}\right)^T \, \cdot \, \mathbf{C}^{-1}_{G} \, \cdot \, \Delta  \boldsymbol{\mathcal{\mu}}^{G}$, $b\equiv\left(\Delta \boldsymbol{\mathcal{\mu}}^{G}\right)^T \, \cdot \, \mathbf{C}^{-1}_{G} \, \cdot \, \boldsymbol{1}$ and $d\equiv\boldsymbol{1}\, \cdot \, \mathbf{C}^{-1}_{G} \, \cdot \, \boldsymbol{1}$.

\subsection{Baryon Acoustic Oscillations}

For BAO we consider data from many different surveys. In general, we can define the $\chi^2$ as
\begin{equation}
\chi^2_{BAO} = \Delta \boldsymbol{\mathcal{F}}^{BAO} \, \cdot \ \mathbf{C}^{-1}_{BAO} \, \cdot \, \Delta  \boldsymbol{\mathcal{F}}^{BAO} \ ,
\end{equation}
with the observables $\mathcal{F}^{BAO}$ which will change depending on which survey is considered.

When using the WiggleZ Dark Energy Survey (at redshifts $0.44$, $0.6$ and $0.73$) \cite{Blake:2012pj}, the observables are the acoustic parameter
\begin{equation}\label{eq:AWiggle}
A(z,\boldsymbol{p}) = 100  \sqrt{\Omega_{m} \, h^2} \frac{D_{V}(z,\boldsymbol{p})}{c \, z} \, ,
\end{equation}
with $h=H_0/100$, and the Alcock-Paczynski distortion parameter
\begin{equation}\label{eq:FWiggle}
F(z,\boldsymbol{p}) = (1+z)  \frac{D_{A}(z,\boldsymbol{p})\, H(z,\boldsymbol{p})}{c} \, ,
\end{equation}
where $D_{A}$ is the angular diameter distance defined in Eq.~(\ref{eq:ang_dist}) and
\begin{equation}
D_{V}(z,\boldsymbol{\theta})=\left[ (1+z)^2 D^{2}_{A}(z,\boldsymbol{\theta}) \frac{c z}{H(z,\boldsymbol{\theta})}\right]^{1/3}
\end{equation}
is the geometric mean of the radial and tangential BAO modes. We stress here that this data set is totally independent of early-time evolution, and for that reason it is included in the late-time data analysis.

We dealing with data from SDSS-III Baryon Oscillation Spectroscopic Survey (BOSS) observations, the situation is different. These BAO data are used for the full data analysis only.

There are many sources and independent ways which have analyzed data from BOSS. In \cite{Alam:2016hwk} the DR$12$ analysis provides:
\begin{equation}
D_{M}(z,\boldsymbol{p}) \frac{r^{fid}_{s}(z_{d},)}{r_{s}(z_{d},\boldsymbol{p})}, \qquad H(z) \frac{r_{s}(z_{d},\boldsymbol{p})}{r^{fid}_{s}(z_{d})} \,,
\end{equation}
where: the comoving distance $D_M$ is
\begin{equation}\label{eq:comovdist}
D_{M}(z,\boldsymbol{p})=\int_{0}^{z} \frac{c\, \mathrm{d}z'}{H(z',\boldsymbol{p})} \; ;
\end{equation}
the sound horizon is
\begin{equation}\label{eq:soundhor}
r_{s}(z,\boldsymbol{p}) = \int^{\infty}_{z} \frac{c_{s}(z')}{H(z',\boldsymbol{p})} \mathrm{d}z'\, ,
\end{equation}
with the sound speed given by
\begin{equation}\label{eq:soundspeed}
c_{s}(z) = \frac{c}{\sqrt{3(1+\overline{R}_{b}\, (1+z)^{-1})}} \; ,
\end{equation}
and the baryon-to-photon density ratio parameters defined as $\overline{R}_{b}= 31500 \Omega_{b} \, h^{2} \left( T_{CMB}/ 2.7 \right)^{-4}$, with $T_{CMB} = 2.726$ K; the sound horizon evaluated at the dragging redshift is $r_{s}(z_{d})$; and the sound horizon calculated for a given fiducial cosmological model (in this case, it is $147.78$ Mpc) is $r^{fid}_{s}(z_{d})$. The dragging redshift is estimated using the analytical approximation provided in  \cite{Eisenstein:1997ik}.

We also consider measurements derived from the DR$12$ and involving the void-galaxy cross-correlation \cite{Nadathur:2019mct}:
\begin{eqnarray}
\frac{D_{A}(z=0.57)}{r_{s}(z_{d})} &=& 9.383 \pm 0.077\,, \\
H(z=0.57) r_{s}(z_{d})  &=& (14.05 \pm 0.14) 10^{3} \, \rm{km\,s^{-1}} \, .
\end{eqnarray}
Instead, from the  extended Baryon Oscillation Spectroscopic Survey (eBOSS) we use the point $D_V(z=1.52)=3843\pm147\,r_s(zd)/r_s^{fid}(z_d)$ Mpc provided by \cite{Ata:2017dya}. Finally, from eBOSS DR14 we have a combination of the Quasar-Lyman $\alpha$ autocorrelation function \cite{lyman} with the cross-correlation measurement \cite{Blomqvist:2019rah}, giving
\begin{eqnarray}
\frac{D_{A}(z=2.34)}{r_{s}(z_{d})} &=& 36.98^{+1.26}_{-1.18}\,, \\
\frac{c}{H(z=2.34) r_{s}(z_{d})}  &=& 9.00^{+0.22}_{-0.22}\, .
\end{eqnarray}

\subsection{Cosmic Microwave Background}

When dealing with CMB data we work with the shift parameters defined in \cite{Wang:2007mza} and derived from the latest \textit{Planck} $2018$ data release \cite{cmb&sn}. The $\chi^2_{CMB}$ is
\begin{equation}
\chi^2_{CMB} = \Delta \boldsymbol{\mathcal{F}}^{CMB} \; \cdot \; \mathbf{C}^{-1}_{CMB} \; \cdot \; \Delta  \boldsymbol{\mathcal{F}}^{CMB} \; ,
\end{equation}
and the vector $\mathcal{F}^{CMB}$ consists of:
\begin{eqnarray}
R(\boldsymbol{p}) &\equiv& \sqrt{\Omega_m H^2_{0}} \frac{r(z_{\ast},\boldsymbol{p})}{c}, \nonumber \\
l_{a}(\boldsymbol{p}) &\equiv& \pi \frac{r(z_{\ast},\boldsymbol{p})}{r_{s}(z_{\ast},\boldsymbol{p})}\,,
\end{eqnarray}
in addition to $\Omega_b\,h^2$. $r_{s}(z_{\ast})$ is the co-moving sound horizon evaluated at the photon-decoupling redshift evaluated using the fitting formula from \cite{Hu:1995en} and $r$ is the co-moving distance at decoupling.

\section{Results and Discussions}
\begin{figure*}[t!]
\centering
\includegraphics[width=0.95\textwidth]{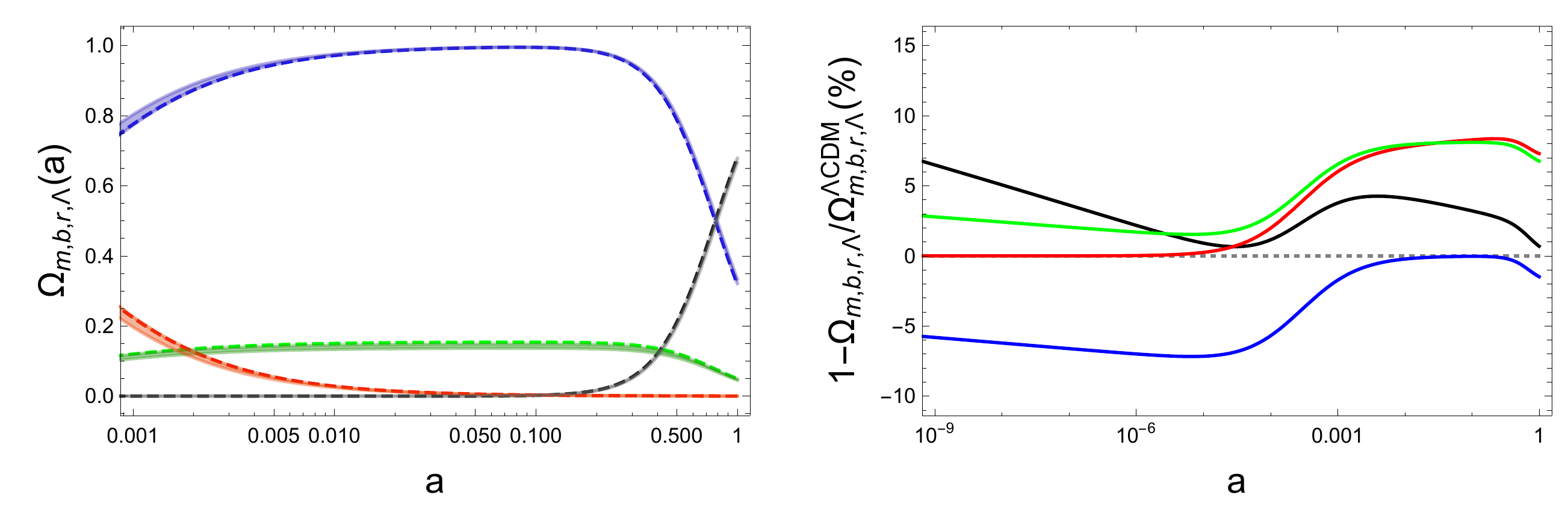}
\caption{Evolution with time of dimensionless density parameters. \textit{(left panel)} Matter (blue), baryons (green), radiation (red) and dark energy (black) for both full Bekenstein entropic model with $\Gamma(t) = \Gamma_0\, H$ ($1\sigma$ confidence levels as shaded regions) and full $\Lambda$CDM (dashed lines). \textit{(right panel)}. Percentage deviation of dimensionless density parameters between full Bekenstein entropic model with $\Gamma(t) = \Gamma_0\, H$ and full $\Lambda$CDM.}
\label{plot:1}
\end{figure*}
Complete results from our statistical analysis are shown in Table~\ref{tab:results1} for the late-time data, and in Table~\ref{tab:results2} for the full data sets. When looking at the results from using only late-time data, we can see how the entropic scenarios are totally indistinguishable from a standard $\Lambda$CDM model, and even among themselves, wherever one looks at the values of the cosmological parameters, or at the Bayes Factors. If we exclude the case with Bekenstein entropy and $\Gamma(t) = \Gamma_0\, H$, which is the only one to be clearly disfavored, and we focus on the main parameters of the entropic models, i.e. $\gamma$ and $\Gamma_0$, we find that they are basically very small, with at least $\log \gamma < -5$ and $\log \Gamma_0 < -4$ at $1\sigma$ confidence levels. Thus, there is a negligible exchange of energy between the bulk and the boundary of the universe. Note that these are only \textit{upper} limits, i.e. these are the maximum values for which the considered models are compatible with present observational data.

When moving to full data, including also early-time observations, as expected, all entropic models have slightly negative value for the Bayes Factors, which means they are slightly disfavored with respect to $\Lambda$CDM. No difference at all is detectable also from the values of the cosmological parameters.

The only successful scenario is exactly the one which was slightly disfavoured in the late-time case, namely the Bekenstein entropic terms with $\Gamma(t) = \Gamma_0\, H$. What is really striking and hitting the attention is the very high and positive value for the Bayes Factor, which states a strong evidence in favor of this scenario with respect to the standard one. While the value of $\Omega_m$ is slightly higher, but still statistically consistent with the result from $\Lambda$CDM, we have a lower value for $\Omega_b$ and a higher one for $H_0$, although both exhibit larger errors with respect to the $\Lambda$CDM case. Note that the shift in $H_0$ is not enough to considerably reduce the Hubble tension \cite{Riess:2020sih}.

Where could this statistical preference come from? Let us start noting that, just for this case, for what concerns the parameters $\gamma$ and $\Gamma_0$, we do retrieve values which are much higher than other cases. While for all other cases we find $\log \gamma < -7$ and $\log \Gamma_0 < -4$, for this high-evidence scenario we do have $\log \gamma \sim -2$ and $\Gamma_0 \sim -0.017$. Note also that, as explained above, this is the only case where $\Gamma_0$ can be negative, and we do not have an upper bound, but a both sided constraint.
{\renewcommand{\tabcolsep}{1.5mm}
{\renewcommand{\arraystretch}{2.}
\begin{table*}
\begin{minipage}{0.99\textwidth}
\caption{Results: late-time}\label{tab:results1}
\centering
\resizebox*{\textwidth}{!}{
\begin{tabular}{c|c|cccccc|ccccccc}
\hline
   & $\Lambda{\rm CDM}$ & \multicolumn{6}{c|}{Bekenstein} & \multicolumn{7}{c}{Tsallis-Cirto} \\
   &                    & \multicolumn{2}{c}{$\Gamma_0$} & $\Gamma_0\,H$ & $\Gamma_0\,H^2$ & \multicolumn{2}{c|}{$\Gamma_0/H$} & \multicolumn{2}{c}{$\Gamma_0$} & \multicolumn{2}{c}{$\Gamma_0\,H$} & $\Gamma_0\,H^2$ & \multicolumn{2}{c}{$\Gamma_0/H$} \\
   &                    & $\Gamma_0>0$ & $\Gamma_0<0$ & & & $\Gamma_0>0$ & $\Gamma_0<0$ & $\Gamma_0>0$ & $\Gamma_0<0$ & $\Gamma_0>0$ & $\Gamma_0<0$ & & $\Gamma_0>0$ & $\Gamma_0<0$ \\
\hline
\hline
$\Omega_{ m}$   & $0.293^{+0.016}_{-0.016}$       & $0.293^{+0.017}_{-0.016}$ & $0.292^{+0.017}_{-0.016}$         & $0.301^{+0.023}_{-0.023}$ & $0.293^{+0.017}_{-0.016}$      & $0.293^{+0.017}_{-0.016}$ & $0.292^{+0.017}_{-0.016}$      & $0.293^{+0.017}_{-0.016}$ & $0.293^{+0.017}_{-0.016}$       & $0.293^{+0.017}_{-0.016}$ & $0.291^{+0.016}_{-0.015}$ & $0.293^{+0.017}_{-0.016}$ & $0.293^{+0.017}_{-0.016}$ & $0.293^{+0.017}_{-0.016}$ \\
$h$             & $0.713^{+0.013}_{-0.013}$       & $0.714^{+0.013}_{-0.013}$ & $0.713^{+0.013}_{-0.013}$         & $0.712^{+0.013}_{-0.013}$ & $0.714^{+0.013}_{-0.013}$               & $0.714^{+0.013}_{-0.013}$ & $0.714^{+0.013}_{-0.013}$            & $0.713^{+0.013}_{-0.013}$ & $0.713^{+0.013}_{-0.012}$              & $0.713^{+0.013}_{-0.013}$ & $0.714^{+0.013}_{-0.013}$ & $0.713^{+0.013}_{-0.012}$    & $0.713^{+0.013}_{-0.013}$ & $0.713^{+0.013}_{-0.013}$ \\
$\log \gamma$   & $-$         & $<-7.26$ & $<-7.03$         & $<-0.713$                 & $<-6.81$        & $<-7.15$ & $<-8.40$     & $<-5.83$ & $<-3.94$   & $<-5.59$ & $-15.8^{+1.7}_{-1.2}$ & $<-5.91$ & $<-5.98$ & $<-7.02$ \\
$\log \Gamma_0$ & $-$          & $<-5.47$ & $<-12.4$          & $-$  & $<-7.29$     & $<-4.39$ & $<-9.43$      & $<-5.57$ & $<-6.74$ & $<-6.68$ & $-2.82^{+0.84}_{-0.85}$ & $<-14.34$ & $<-4.29$ & $-29.7^{+15.3}_{-15.1}$ \\
$\Gamma_0$      & $-$   & $-$ & $-$         & $-0.17^{+0.20}_{-0.28}$   & $-$ & $-$ & $-$ & $-$ & $-$ & $-$ & $-$ & $-$ & $-$ \\
\hline
$\chi^2$                        & $1094.17$     & $1093.74$ & $1093.92$     & $1093.79$ & $1093.83$        & $1093.75$ & $1093.92$     & $1093.74$ & $1093.90$ & $1093.81$ & $1094.17$ & $1093.84$ & $1093.72$ & $1093.89$ \\
$\mathcal{B}_{j}^{i}$           & $\mathit{1}$  & $1.01^{+0.02}_{-0.02}$ & $0.99^{+0.03}_{-0.02}$  & $0.50^{+0.01}_{-0.01}$  & $1.02^{+0.02}_{-0.02}$   & $1.01^{+0.03}_{-0.03}$ & $0.99^{+0.02}_{-0.03}$ & $1.02^{+0.03}_{-0.02}$ & $1.00^{+0.02}_{-0.03}$  & $1.01^{+0.02}_{-0.03}$ & $0.98^{+0.02}_{-0.02}$ & $1.02^{+0.02}_{-0.02}$ & $1.02^{+0.03}_{-0.02}$ & $1.00^{+0.03}_{-0.03}$ \\
$\ln \mathcal{B}_{j}^{i}$       & $\mathit{0}$  & $0.009^{+0.023}_{-0.021}$  & $-0.005^{+0.029}_{-0.024}$ & $-0.68^{+0.02}_{-0.02}$ & $0.02^{+0.02}_{-0.02}$  & $0.01^{+0.03}_{-0.03}$  & $-0.007^{+0.019}_{-0.027}$ & $0.02^{+0.03}_{-0.02}$ & $0.002^{+0.024}_{-0.027}$ & $0.007^{+0.021}_{-0.031}$ & $-0.02^{+0.03}_{-0.03}$ & $0.02^{+0.02}_{-0.02}$ & $0.01^{+0.03}_{-0.02}$ & $-0.004^{+0.026}_{-0.027}$ \\
\hline
\hline
\end{tabular}}
\end{minipage}
\end{table*}}}

{\renewcommand{\tabcolsep}{1.5mm}
{\renewcommand{\arraystretch}{2.}
\begin{table*}
\begin{minipage}{0.99\textwidth}
\caption{Results: full}\label{tab:results2}
\centering
\resizebox*{\textwidth}{!}{
\begin{tabular}{c|c|cccccc|ccccccc}
\hline
   & $\Lambda{\rm CDM}$ & \multicolumn{6}{c|}{Bekenstein} & \multicolumn{7}{c}{Tsallis-Cirto} \\
   &                    & \multicolumn{2}{c}{$\Gamma_0$} & $\Gamma_0\,H$ & $\Gamma_0\,H^2$ & \multicolumn{2}{c|}{$\Gamma_0/H$} & \multicolumn{2}{c}{$\Gamma_0$} & \multicolumn{2}{c}{$\Gamma_0\,H$} & $\Gamma_0\,H^2$ & \multicolumn{2}{c}{$\Gamma_0/H$} \\
   &                    & $\Gamma_0>0$ & $\Gamma_0<0$ & & & $\Gamma_0>0$ & $\Gamma_0<0$ & $\Gamma_0>0$ & $\Gamma_0<0$ & $\Gamma_0>0$ & $\Gamma_0<0$ & & $\Gamma_0>0$ & $\Gamma_0<0$ \\
\hline
\hline
$\Omega_{ m}$   & $0.319^{+0.005}_{-0.005}$       & $0.319^{+0.005}_{-0.005}$ & $0.319^{+0.005}_{-0.005}$     & $0.325^{+0.005}_{-0.005}$  & $0.319^{+0.005}_{-0.005}$                & $0.319^{+0.005}_{-0.005}$ & $0.320^{+0.004}_{-0.004}$               & $0.319^{+0.005}_{-0.005}$ & $0.320^{+0.005}_{-0.005}$               & $0.319^{+0.005}_{-0.005}$ & $0.319^{+0.005}_{-0.005}$ & $0.319^{+0.005}_{-0.005}$ & $0.319^{+0.005}_{-0.005}$ & $0.319^{+0.005}_{-0.005}$ \\
$\Omega_{ b}$   & $0.0494^{+0.0004}_{-0.0004}$    & $0.0494^{+0.0004}_{-0.0004}$ & $0.0494^{+0.0004}_{-0.0004}$ &          $0.0466^{+0.0012}_{-0.0012}$ & $0.0494^{+0.0004}_{-0.0004}$             & $0.0494^{+0.0004}_{-0.0004}$ & $0.0494^{+0.0004}_{-0.0004}$             & $0.0494^{+0.0004}_{-0.0004}$ & $0.0494^{+0.0004}_{-0.0004}$          & $0.0494^{+0.0004}_{-0.0004}$  & $0.0494^{+0.0004}_{-0.0004}$ & $0.0494^{+0.0004}_{-0.0004}$         & $0.0494^{+0.0004}_{-0.0004}$ & $0.0494^{+0.0004}_{-0.0004}$ \\
$h$             & $0.673^{+0.003}_{-0.003}$       & $0.672^{+0.004}_{-0.003}$ & $0.673^{+0.003}_{-0.003}$       & $0.694^{+0.010}_{-0.009}$  & $0.673^{+0.003}_{-0.003}$               & $0.672^{+0.003}_{-0.003}$ & $0.672^{+0.004}_{-0.003}$                & $0.672^{+0.003}_{-0.003}$ & $0.673^{+0.004}_{-0.003}$         & $0.673^{+0.003}_{-0.003}$ & $0.673^{+0.004}_{-0.003}$ & $0.672^{+0.003}_{-0.003}$              & $0.672^{+0.003}_{-0.003}$ & $0.672^{+0.003}_{-0.003}$ \\
$\log \gamma$   & $-$ & $<-8.84$ & $<-9.01$       & $<-2.23$ & $<-13.40$           & $<-10.99$ & $<-10.91$     & $<-6.72$ & $<-7.17$          & $<-7.07$ & $<-7.98$ & $<-6.78$ & $<-9.26$ & $<-9.25$ \\
$\log \Gamma_0$ & $-$ & $<-6.86$ & $<-19.9$  & $-$                        & $-31.5^{+2.6}_{-2.7}$ & $<-5.96$ & $<-10.03$ & $<-8.85$ & $<-9.38$ & $<-8.76$ & $<-8.86$ & $<-31.09$ & $<-3.78$ & $<-2.14$ \\
$\Gamma_0$      & $-$ & $-$ & $-$ & $-0.017^{+0.008}_{-0.012}$   & $-$ & $-$ & $-$ & $-$ & $-$ & $-$ & $-$ & $-$ & $-$\\
\hline
$\chi^2$                       & $1124.25$  & $1124.44$ & $1121.01$ & $1114.34$ & $1124.45$ & $1124.44$ & $1121.69$ & $1124.44$ & $1121.07$ & $1124.44$ & $1191.91$ & $1124.44$ & $1124.44$ & $1124.44$ \\
$\mathcal{B}_{j}^{i}$          & $\mathit{1}$  & $0.90^{+0.03}_{-0.03}$ & $1.09^{+0.03}_{-0.03}$ & $61.7^{+1.6}_{-1.8}$ & $0.93^{+0.03}_{-0.02}$ & $0.89^{+0.03}_{-0.03}$ & $1.15^{+0.03}_{-0.03}$ & $0.90^{+0.03}_{-0.02}$ & $1.49^{+0.05}_{-0.04}$ & $0.89^{+0.02}_{-0.03}$ & $1.11^{+0.04}_{-0.02}$ & $0.91^{+0.02}_{-0.03}$ & $0.91^{+0.02}_{-0.02}$ & $0.90^{+0.03}_{-0.02}$ \\
$\ln \mathcal{B}_{j}^{i}$      & $\mathit{0}$  & $-0.10^{+0.03}_{-0.03}$ & $0.08^{+0.03}_{-0.03}$ & $4.12^{+0.03}_{-0.03}$ & $-0.08^{+0.03}_{-0.03}$ & $-0.11^{+0.04}_{-0.03}$ & $0.14^{+0.03}_{-0.03}$ & $-0.11^{+0.04}_{-0.02}$ & $-0.11^{+0.03}_{-0.03}$ & $0.40^{+0.03}_{-0.03}$ & $0.10^{+0.04}_{-0.02}$ & $-0.10^{+0.03}_{-0.03}$ & $-0.10^{+0.03}_{-0.03}$ & $-0.11^{+0.03}_{-0.02}$ \\
\hline
\hline
\end{tabular}}
\end{minipage}
\end{table*}}}

Moreover, as described in previous sections, in this scenario we modify the time behavior of the cosmological fluids (except the cosmological constant) by the factor $(1-\gamma)(1-\Gamma_0/3)$. In $\Lambda$CDM this is equal to one, while here from our MCMCs we do infer the value $1.0014^{+0.0006}_{-0.0006}$, so that the standard value is excluded at $2\sigma$.

Eventually, more insights can be derived from plotting the behavior of the dimensionless density parameters,
\begin{equation}\label{eq:dimensionless_dens}
\Omega_{m,b,r}(a) = \frac{H^{2}_0}{H^{2}(a)} \frac{\Omega_{m,b,r} a^{-3(1+w_{m,b,r})(1-\gamma)(1-\Gamma_0/3)}}{1-\gamma} \,,
\end{equation}
\begin{equation}\label{eq:dimensionless_dens_DE}
\Omega_{\Lambda}(a) = \frac{H^{2}_0}{H^{2}(a)} \frac{1-\Omega_m-\Omega_r-\gamma}{1-\gamma} \,.
\end{equation}
We show them in Fig.~\ref{plot:1}. In the left panel, we show the time dependence of matter (blue), baryons (green), radiation (red) and dark energy component (black), with $1\sigma$ confidence levels as shaded regions for the Bekenstein entropic model, against the $\Lambda$CDM ones, given by dashed line with the same colors. While we can detect visually a statistically light deviation between the two cases, in the right panel we plot percentage deviations between the $\Lambda$CDM quantities and the corresponding entropic models ones. There it is more clear that such model behaves as a \textit{light} early dark energy model, with $\gtrsim5\%$ more dark energy at very early times; a fast decrease till $a\sim 10^{-4}$; a new rise till recombination physics, where it requires $\sim 3-4\%$ more dark energy, with a peak of $5\%$ again right after recombination; and then a slow decrease till present time. Seemingly, this small excess of dark energy, accompanied by a smaller amount of matter (up to $6\%$ before recombination) can be the origin of such improvement in the fit.

It is interesting to note that the value of $\gamma$ is considered to be of order one theoretically in the literature \cite{Easson:2010av,Komatsu:2013qia}. In \cite{Dabrowski:2015tia,Gohar:2017xsr}, $\gamma$ has been constrained by using entropic force models combined the variation of fundamental constants and it results to be $\sim 10^{-4}-10^{-2}$, which is completely different from the theoretical expectations. In literature \cite{Gomez-Valent:2014fda,Sola:2007sv,Gomez-Valent:2017idt,Koivisto:2010tb,Gomez-Valent:2018nib,Sola:2017jbl,Grande:2010vg,Fabris:2006gt,Shapiro:2003ui,AgudeloRuiz:2019nnm} one can find ``similar'' parameters, i.e. related to energy exchange phenomena, although they do not point to the system ``boundary-bulk'' but to energy exchanges among cosmological fluids. Thus, any comparison with our analysis would not be appropriate.

The $\Gamma(t)$ function has been extensively studied before, but in contexts which are totally different from ours, as bulk viscous cosmology and the creation of dark matter cosmological models \cite{Abramo:1996ip,Zimdahl:1996ka,Pan:2018ibu,Paliathanasis:2016dhu,Pan:2016jli,Nunes:2015rea,Nunes:2016aup,Ramos:2014dba,Santos:2014kia}. In \cite{Pan:2016jli}, a phenomenological $\Gamma(t)=-\Gamma_0+3 H(t)+\Gamma^{2}_0/H(t)$ function has been constrained, with $\Gamma_0 \sim 10^{-7}$ (see Pag.~10 of \cite{Pan:2016jli}). Clearly, such a value is quite consistent with our results. In \cite{Nunes:2015rea}, three different $ \Gamma(t)$ functions have been studied in the context of matter creation cosmologies. In particular, the model with $\Gamma(t)=3\beta H$, given $\beta\ge0$, provides $\beta=0.0729^{+0.035}_{-0.034}$ for older SNeIa+GRB+BAO+H(z) data sets, and CMB is not included. Note that, depending on the entropy form, our estimations are or at least four order of magnitudes smaller (for Tsallis-Cirto models), or even the sign of $\Gamma_0$ is totally different (for Bekenstein entropy we left the sign totally free and negative values have higher statistical preference). Although, let us stress again that these constraints have not been derived in the context of entropic cosmology models so that any comparison is not straightforward.

\section{Conclusions}
In this work we have investigated entropic force cosmological models with matter creation and energy exchange between the bulk and the boundary of the universe. For this purpose, we have used the Bekenstein entropy, the Tsallis Cirto entropy and the quartic entropy to model three different entropic cosmological scenarios, and a general phenomenological functional form for the matter creation rate. By using the latest observational data sets, we have shown that the energy exchange between the bulk and the boundary is negligible as well as the matter creation rate.

However, most of the entropic cosmological models considered here are basically indistinguishable from a standard $\Lambda$CDM cosmological model. Only in one case, the Bekenstein entropic model with $\Gamma(t)=\Gamma_0H$, the model resulted to be statistically highly favourable over $\Lambda$CDM, with the parameters leading energy exchange between the bulk and the boundary being $\sim 10^{-2}$, namely, quite larger than other cases.

\bibliography{ref}{}
\bibliographystyle{apsrev4-1}

\end{document}